\documentclass{article}

\usepackage[letterpaper,top=2cm,bottom=2cm,left=2cm,right=2cm,marginparwidth=1.75cm]{geometry}
\usepackage{multicol}
\usepackage{amsmath}
\usepackage{graphicx}
\usepackage{url}
\usepackage{natbib}

\usepackage{multirow}  
\usepackage{xcolor}
\usepackage{lscape}
\usepackage{textcomp}
\usepackage{txfonts}
\usepackage{amsmath,bm}
\usepackage{color}
\usepackage{mathtools}
\usepackage{hyperref}
\usepackage{sidecap}

\usepackage{authblk}

\title{Exoplanets in reflected starlight with dual-field interferometry and a fifth Unit Telescope at VLTI}
\author[1]{Ó. Carrión-González}
\author[2,3]{S. Lacour}
\author[2]{M. Nowak}
\affil[1]{Max-Planck-Institut f\"ur Astronomie, K\"onigstuhl 17, 69117 Heidelberg, Germany}
\affil[2]{LIRA, Observatoire de Paris, PSL, CNRS, Sorbonne Universit\'e, Universit\'e de Paris, 5 place Janssen, 92195 Meudon, France}
\affil[3]{European Southern Observatory, Karl-Schwarzschild-Stra\ss{}e 2, 85748 Garching, Germany}
\date{}                     
\begin{document}
\maketitle

\clearpage
\begin{abstract}
In this white paper, we propose an upgrade to the Very Large Telescope
Interferometer (VLTI) consisting of the addition of a new 8\,m Unit
Telescope (UT5). The primary goal of this upgrade is to optimise the VLTI
for exoplanet detection by creating four additional baselines of
approximately 200\,m oriented toward the north-west. The inclusion of
this telescope would reduce the inner working angle and improve the
achievable contrast of the VLTI, thereby enabling the detection of
mature exoplanets in reflected light.
\end{abstract}

\section{A journey toward direct imaging of mature exoplanets} \label{sec:intro}

In about 20 years since the first exoplanet detections in direct imaging \citep{chauvinGiantPlanetCandidate2004, maroisDirectImagingMultiple2008}, this technique has become a key for the characterisation of exoplanets. 
The initial direct detections were self-luminous giant planets in very early stages of formation, given their favourable planet-to-star contrast levels.
However, with the improvements in instrumental sensitivity, e.g. with the James Webb Space Telescope (JWST) coronagraphic capabilities \citep{boccalettietal2022, kammereretal2022SPIE12180E..3NK}, we are now capable of directly imaging a handful of temperate super-Jupiters in the mid-infrared \citep{matthewsetal2024Natur.633..789M}.
In 2027, the Coronagraph Instrument on the Nancy Grace Roman Space Telescope \citep[hereon, Roman;][]{spergeletal2015} will provide the first direct-imaging observations of cold exoplanets in reflected starlight.
This will open the path to studying the population of cold and temperate long-period exoplanets in the neighbourhood of the Solar System.
In the coming decades, the direct imaging technique has been identified as a priority for the characterisation of temperate exoplanets by both the US Astro 2020 Decadal Survey, and by ESA’s Voyage 2050 Senior Committee report.
This has resulted, respectively, in the Habitable Worlds Observatory --based on the LUVOIR \citep{luvoirteam2018} and HabEx \citep{mennessonetal2016, gaudietal2018} studies--, and the mid-IR Large Interferometer For Exoplanets \citep[LIFE][]{2022A&A...664A..21Q} projects.

\begin{figure}[h]
   \centering
   \includegraphics[width=15.5cm]{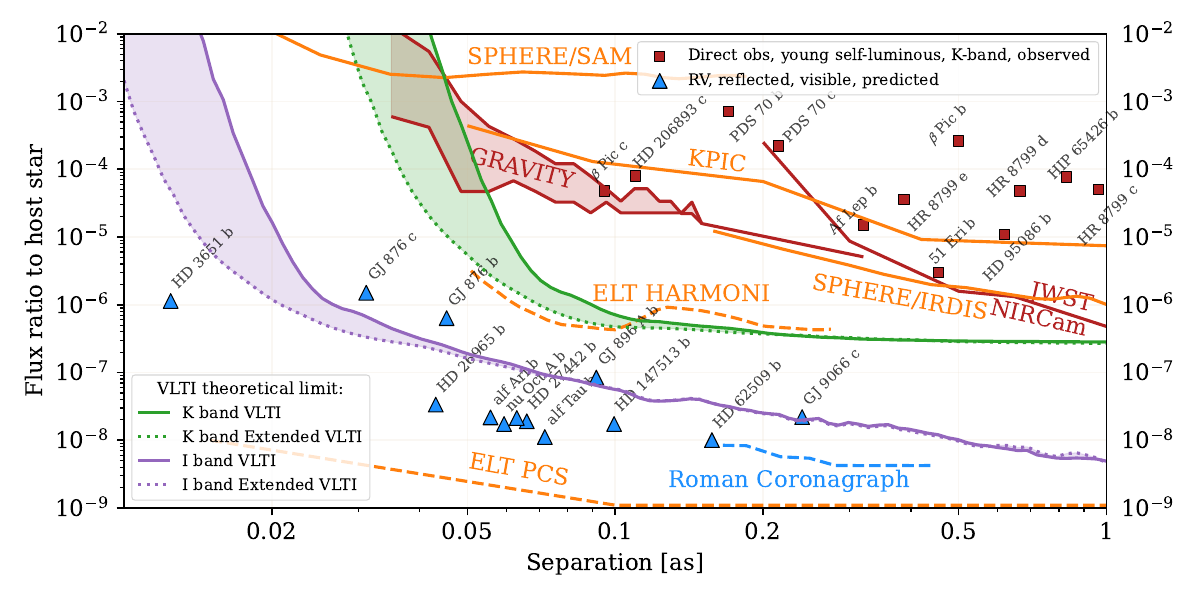}
   \caption{Overview of current and predicted contrast limits for present and future direct-imaging instruments. Red squares represent directly detected exoplanets, and blue triangles show the contrast predictions for radial-velocity exoplanets. Figure from \citet{lacouretal2025A&A...694A.277L}, see Fig. 11 therein.}
\label{fig:final_contrast}
\end{figure}

On the ground, work is ongoing for the design of the Planetary Camera and Spectrograph (PCS), a future instrument at ESO's Extremely Large Telescope (ELT) capable of directly imaging temperate low-mass exoplanets in reflected starlight \citep{kasperetal2021}.
Before PCS, projects to directly image exoplanets in reflected light from the ground include the proposed RISTRETTO high-resolution spectrograph \cite{bugattietal2025A&A...702A.230B} at the Very Large Telescope (VLT), or the expansion of the GRAVITY instrument at the Very Large Telescope Interferometer (VLTI) towards shorter wavelengths in the J band \citep{lacouretal2025A&A...694A.277L}, which has been funded and is undergoing the design phase \footnote{ERC PLANETES (PI: Lacour): \url{https://cordis.europa.eu/project/id/101142746/fr}}.
So far, the GRAVITY instrument \citep{gravitycollaborationFirstLightGRAVITY2017} has pioneered dual-field interferometry observations, and was the first interferometer to directly detect an exoplanet \citep{gravitycollaborationFirstDirectDetection2019}. 
GRAVITY also provided the first direct detection of $\beta$\,Pictoris\,c \citep{nowakDirectConfirmationRadialvelocity2020} and HD\,206893\,c \citep{hinkleyDirectDiscoveryInner2023}, two young exoplanets orbiting at small angular separations of their host stars, and originally discovered in radial velocity.
These two exoplanets have not been directly imaged with any other instrument to date, suggesting that dual-field interferometry performs significantly well at small angular separations. 
With the commissioning of GRAVITY+, extreme adaptive optics (ExAO) will be available at all four Unit Telescopes (UTs) of the VLTI.
This, coupled with dual-field interferometry \citep{2022Msngr.189...17A}, will further increase this potential for small-separation measurements, and optimise the VLTI for high-contrast observations \citep{pourreetal2024A&A...686A.258P}.

\section{Unique questions answered with an extended VLTI} \label{sec:science}

In \citet{lacouretal2025A&A...694A.277L} we studied the population of about 450 known exoplanets within 30~pc reported in the NASA Exoplanet Archive \citep{2013PASP..125..989A}.
We used the methodology introduced in \citet{carriongonzalezetal2021a}, and computed 1000 orbital realisations for each of these 450 planets - varying the values of the orbital parameters within the uncertainties reported in the NASA Archive.
Throughout each orbit, we computed the variation of angular separation of the planet, and its planet-to-star contrast ratio in reflected light at band I ($\sim$800~nm).
We assumed Lambertian scattering and a geometric albedo of 0.3 for all planets.
We then computed the integration time required to detect the planet using the instrumental models from \citet{lacouretal2025A&A...694A.277L}, assuming a detection threshold of $\mathrm{S/N}$=3.

\begin{SCfigure}[][h]
  \centering
  \includegraphics[width=0.5\textwidth]{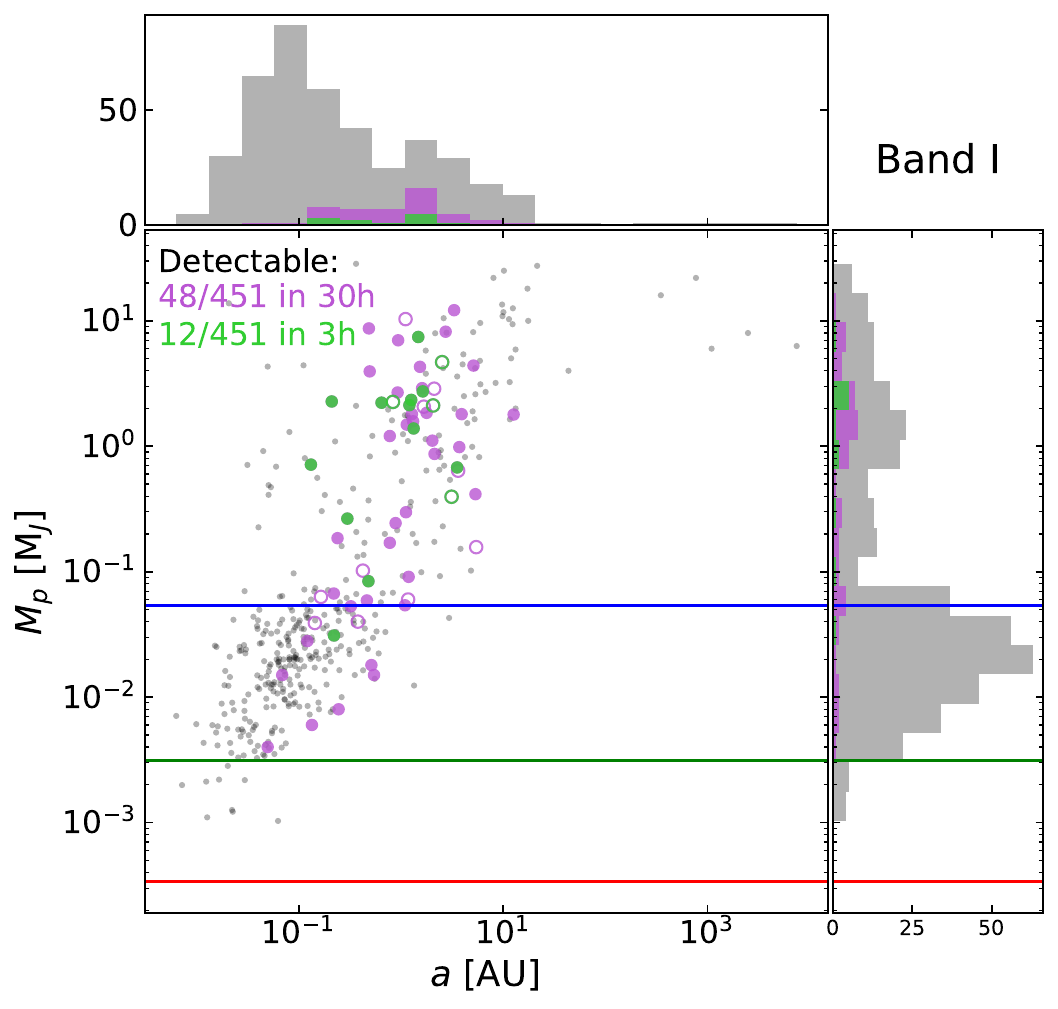}
	\caption{Detectable exoplanets with a proposed fifth UT at the VLTI, with 220~m baselines and operating in the I band. Purple (resp. green) circles and histograms show the detectable planets in 30~h (resp. 3~h) of integration time, according to the instrumental models developed in \citet{lacouretal2025A&A...694A.277L}.
    Grey dots and histograms show the ensemble of known exoplanets within 30~pc.
    Horizontal lines indicate the masses of Neptune (blue), Earth (green) and Mars (red). Figure from \citet{lacouretal2025A&A...694A.277L}, see Fig. 9 therein for details.}
\label{fig:results_MpVSd}
\end{SCfigure}

We found that up to 48 of the 451 exoplanets within 30~pc would be detectable in band I with this extended VLTI with a fifth UT (Fig. \ref{fig:results_MpVSd}).
All of these targets require less than 30~h of integration time for detection, with 12 of them needing less than 3~h.
Fig. \ref{fig:results_MpVSd} shows that this instrument would explore a unique space of parameters, reaching exoplanets down to the rocky regime. 
Proxima Centauri b, for instance, would be detectable thanks to the improved capabilities of the interferometer in terms of inner working angle (IWA), as also shown in Fig. \ref{fig:contrast_VLTI}.
Several of the exoplanets detectable in Fig. \ref{fig:results_MpVSd} are also potential targets of the Roman coronagraph, PCS, HWO or LIFE \citep{kasperetal2021, carriongonzalezetal2021a, carriongonzalezetal2023A&A...678A..96C}.
This opens the path for a multi-facility, multi-wavelength observation of those targets, which will be a key for their optical and atmospheric characterization.
Indeed, in the case of reflected-light observations of exoplanets, retrieval studies on simulated observations have shown that an accurate orbital characterization is required to constrain the atmospheric properties of the planets \citep{nayaketal2017PASP..129c4401N, carriongonzalezetal2021b, salvadoretal2024ApJ...969L..22S, damianoetal2025AJ....169...97D}.
Upcoming and future direct imaging facilities observing in reflected light --in particular those in space-- will have limited capabilities and time availability for orbital characterization and follow-up observations.
These facilities will greatly benefit from a complementary instrument, such as the proposed VLTI extension, for the orbital characterization of nearby exoplanetary systems -- a task in which the GRAVITY instrument has already showcased excellent performance \citep[e.g.][]{lacourMassPictorisPictoris2021, wangConstrainingNaturePDS2021}.

\section{Physical implementation of the fifth UT}

\begin{figure}
   \begin{center}
      \includegraphics[width=0.8\linewidth]{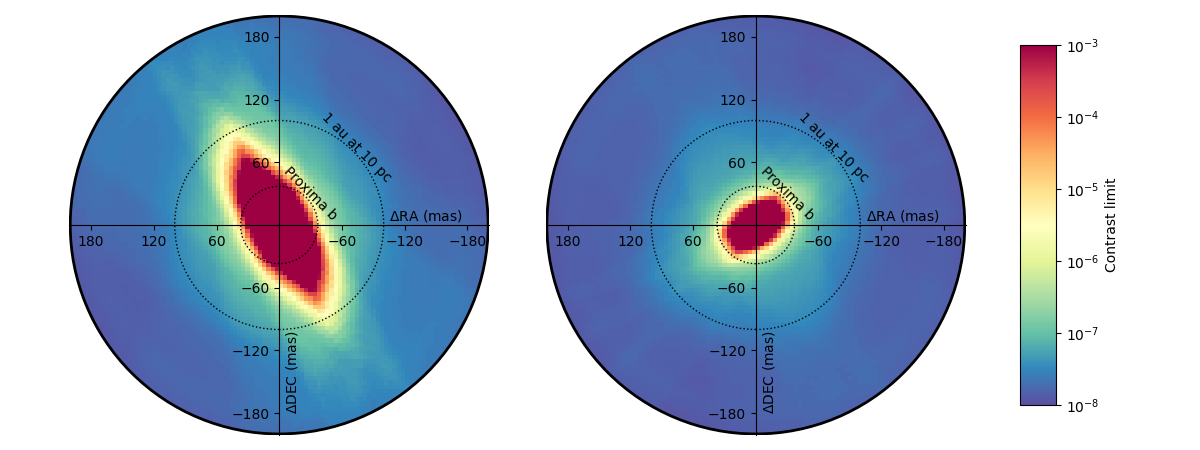}
      \caption{Contrast limit of a dual-field interferometer using the current four-UT configuration of the VLTI (left panel), compared to the proposed five-UT configuration (right panel). Figure from \citet{lacouretal2025A&A...694A.277L}.}
      \label{fig:contrast_VLTI}
   \end{center}
\end{figure}
Our proposal for adding a fifth UT would enable the VLTI to address the unique scientific questions  discussed above, without extensive change to the VLTI architecture taking into account the technical challenges: need to both enhance the UV coverage in the North-West direction, and to increase the baseline lengths for maximising the inner working angle.

This new telescope could be constructed to the south of UT4, across the road. It would be aligned with the AT J band arm, facilitating the propagation of the beam towards the delay line tunnels. The required delay line for a baseline of approximately 200~m could be achieved by serially utilising the two remaining unused delay lines of the VLTI. An aerial view of this proposed setup is available in \citet[][Fig. 8 therein]{lacouretal2025A&A...694A.277L}.

In terms of UV coverage, the four new baselines would ideally complement the existing frequency coverage. With these four new baselines, the IWA would also reach optimal levels, approaching the theoretical limit of $\lambda/D$. Additionally, this configuration would enhance the contrast limit at many position angles, improving the detectability of exoplanets. This setup would enable the detection of an exoplanet at a 1 AU distance from a star located 10 parsecs away across all position angles. This setup would also allow for the detection of Proxima Centauri b at favourable orbital positions. Figs. \ref{fig:final_contrast} and \ref{fig:contrast_VLTI} show the enhancements provided by these additional capabilities. See \citet{lacouretal2025A&A...694A.277L} for more details. 

In summary, the addition of this fifth UT would result in:
\begin{itemize}
    \item Longest maximum baseline increased from current 100~m to 220~m.
    \item More thorough coverage of the UV plane, given current baseline limitations in the North-West direction (max. baseline in that direction is currently 50~m).
    \item Reduced IWA, enabling the detection of exoplanets at closer separations.
    \item Increased astrometric capabilities for the orbital characterization of exoplanets.
\end{itemize}

\begin{multicols}{2}
\footnotesize
\setlength{\bibsep}{0pt}
\bibliographystyle{aa}
\bibliography{sample}
\end{multicols}

\end{document}